\documentclass[aps,pra,onecolumn,showpacs,amsmath,superscriptaddress]{revtex4}
\usepackage{xspace,epsfig,amsfonts,amssymb}
\usepackage{graphics,amstext,color}

\newcommand{\beq}{\begin{equation}} 
\newcommand{\eeq}{\end{equation}} 

\newcommand{\ba}{\begin{array}} 
\newcommand{\ea}{\end{array}} 

\def\beqa{\begin{eqnarray}}
\def\eeqa{\end{eqnarray}}

\newcommand{\ket}[1]{\vert \mkern1mu {#1} \mkern1mu \rangle} 
\newcommand{\bracket}[2]{\langle \mkern1mu {#1} \mkern1mu \vert \mkern1mu {#2} \mkern1mu \rangle} 

\newcommand{\arrow}[1]{%
 Ê\savebox0{\hbox{$\scriptstyle{#1}$}}\dimen0\wd0\advance\dimen0 0.7em
 Ê\stackrel{#1}{\hbox to \dimen0 {\hss\rightarrowfill\hss}}}

\begin{document}

\baselineskip 13pt

\title{Improving the Control Strategy in two-way deterministic cryptographic protocols}

\author{Anita Eusebi}
\email{anita.eusebi(at)unicam.it}

\author{Stefano Mancini}
\email{stefano.mancini(at)unicam.it}

\affiliation{School of Science and Technology, University of Camerino, 
I-62032 Camerino, Italy, EU}

\begin{abstract} 
We introduce a new control strategy on a two-way deterministic cryptographic 
scheme, which relies on a suitable unitary transformation rather than quantum 
measurement.
The study is developed for $d$-ary alphabets and the particular choice of the 
transformation works when $d$ is an odd prime power.
It leads to an improvement of the protocol security, which we prove to increase 
with the alphabet order $d$. 
\end{abstract}


\maketitle


\section{Introduction}

The pioneering protocol for Quantum Key Distribution (QKD) is known to be the BB84 
\cite{BB84}. This allows two remote parties (Alice and Bob) to share a secret key 
by a \emph{unidirectional} use of a quantum channel. 
It has a \emph{probabilistic} character, that is, on each use of quantum channel, 
the sender (Alice) is not sure that the encoded symbol will be correctly decoded 
by the receiver (Bob).

In the last decade a new generation of protocols has been introduced realizing 
QKD processes in a \emph{deterministic} way \cite{BEKW02, BF02, CL04, LM05, LM05new}. 
In this case Alice is sure about the fact that Bob will exactly decode the symbol 
she has encoded.
Another important feature of the protocols defined in \cite{BF02, CL04, LM05, LM05new} 
is the \emph{bidirectional} use of the quantum channel.

As much as like extensions of BB84 to larger alphabets have been developed 
\cite{BT99, CBKG01}, there is a number of works extending the deterministic 
protocols proposed in \cite{CL04, LM05} to higher dimensions, in particular for a 
tri-dimensional alphabet \cite{SLW06, SW07}, for a continuous infinite-dimensional 
alphabet \cite{PMBL06} and for $d$-ary alphabets with $d$ prime power dimension 
\cite{EM09}.

In all these cases the security of the protocol is guaranteed by a control process,
which amounts to perform quantum measurements by Alice and the subsequent comparison 
on the public channel of bases used by Alice and Bob.

In this paper, by considering the general two-way deterministic protocol proposed 
in \cite{EM09}, we suggest a new strategy for the control process.
More precisely, we show that it can be realized by a suitable unitary transformation
as well.

Moreover, we study the same powerful eavesdropping attack as in \cite{EM09} on the 
forward and backward path of the quantum channel and we obtain an improvement 
of the security performance. 
In particular we show that the security of the protocol increases in terms of 
the alphabet order $d$.

Finally, we also address the issue of Quantum Direct Communication (QDC) 
\cite{BF02, DLL03, DL04, CL04dir} and see that in this case the optimal dimension 
is $d=3$.

Our protocol is based on Mutually Unbiased Bases (MUB) \cite{I81, WF89, BBRV01, KR03, D05}, 
so it generally works for prime power dimensions $d$. 
But our new strategy of control is valid for only odd prime powers, then we limite our 
work to this case.


\section{The protocol}


Let us consider a qudit, i.e., a $d$-dimensional quantum system, and indicate 
with $\mathcal{H}_{d}$ the associated Hilbert space.
A set of orthonormal bases in $\mathcal{H}_{d}$ is called a set of 
\emph{Mutually Unbiased Bases} (MUB) if the absolute value of the inner product of 
any two vectors from different bases is $1/\sqrt{d}$ (the MUBness condition) 
\cite{EM09,I81, WF89, BBRV01, KR03}. 

At the present, no example of maximal set is known if the Hilbert space dimension 
is a composite number, otherwise it is known that there exists a maximal set of $d+1$ 
MUB in Hilbert spaces of prime power dimension $d=p^m$ with $p$ a prime number and 
$m$ positive integer \cite{I81, WF89, BBRV01, KR03}.

Here, we focus on this case and from now we denote the $d+1$ MUB of $\mathcal{H}_{d}$ 
by $\ket{v_t^{k}}$, with $k = 0, 1, \ldots, d$ and $t = 0, 1, \ldots, d-1$ 
labelling the basis and the vector in it respectively.

Let us denote $\omega$ the $p$-th root of unity $e^{i2\pi /p}$.
Hence, we choose $\{\ket{v_t^0}\}_{t=0, \ldots, d-1}$ as the computational basis 
and use the explicit formula given in \cite{D05} for MUB's vectors to express the 
vectors of any other basis in the following compact way: 
\vskip-18pt
	\beq \label{ket_d}
\ket{v^k_t} = {1 \over \sqrt d} 
\sum_{q=0}^{d-1}
\omega^{\ominus q \odot t}
(\omega^{(k-1) \odot q \odot q})^{\frac{1}{2}}
\ket{v^0_q}
\quad \textnormal{ \ with \ }  
k = 1, \ldots, d 
\textnormal{ \ and \ } 
t = 0, 1, \ldots, d-1 \, .
	\eeq

This expression satisfies the MUBness condition for $d$ any prime power, both 
even and odd (see Appendix in \cite{EM09} for the even case).
However, in the following we make use of (\ref{ket_d}) only in the case of odd 
prime power dimensions. 

In this context, we deal with the Galois field $G=\mathbb{F}(p^m)$ of $d$ elements,
according to its mathematical properties.
Notice that finite fields with $d$ elements exist if and only if $d$ is a prime 
power. In particular, we denote by $\oplus$, $\odot$ and $\ominus$ respectively the 
addition, the multiplication and the subtraction in the field $G$. 
Usually, an element of $G$ is represented by a $m$-tuple $(g_0, g_1, \ldots, g_{m-1})$ 
of integers modulo $p$. According to this representation, $\oplus$ corresponds to the 
componentwise addition modulo $p$ (this is a direct consequence of the fact that, for 
all finite fields of $d=p^m$ elements, the characteristics of the field is exactly 
the prime number $p$).

Following \cite{D05}, we identify $G$ with $\{0, 1, \ldots, d-1\}$, paying attention 
to distinguish the operations in the field from the usual ones.
Namely, we identify $(g_0, g_1, \ldots, g_{m-1})$ with the integer 
$g=\sum_{n=0}^{m-1}g_{n}p^{n}$.
This allows us to consider the vector label $t$ in $\ket{v_t^{k}}$ as an element 
of $G$ and to write $\omega^{g}$ with $g \in G$ (notice that in this way we have
$\omega^{g} = \omega^{g_{0}}$).

As in \cite{EM09}, we consider Bob sending to Alice a qudit state randomly chosen 
from the set $\{\ket{v_{t}^{k}}\}^{k = 1, \ldots, d}_{t = 0, \ldots, d-1}$ of MUB. 
Then, whatever is the state, Alice has to encode a symbol belonging to a $d$-ary 
alphabet $A = \{0, \ldots, d-1\}$ in such a way that Bob will be able to unambiguously 
decode it (notice that the alphabet $A$ can be identified with the Galois field $G$). 
Besides encoding Alice has to perform a control process
to guarantee the security of the protocol.


\subsection{Encoding process}

As in \cite{EM09}, we consider the unitary transformations $V_0^a$ for $a \in A$, 
defined by
	\beq
V^a_0 \, \ket{v^0_t} 
= \omega^{t \odot a} \, \ket{v^0_t} \, , 
	\eeq
which can be regarded as the generalized Pauli $Z$ operators. 

Such operator $V^a_0$ realizes the same shift on all the bases but the 
computational one, that is for $k > 0$:
	\beq
V^a_0 \, \ket{v^k_t}
= {1 \over \sqrt d} \sum_{q=0}^{d-1}
\omega^{\ominus q \odot (t \ominus a)}
(\omega^{(k-1) \odot q \odot q})^{\frac{1}{2}} \ket{v^0_q}
= \ket{v^k_{t \ominus a}} \, .
	\eeq

Then, Alice encoding operation will be the shift operation realized by this 
operator $V_0^a$ for $a \in A$. 
In such a case, Bob receiving back the state $\ket{v^k_{t \ominus a}}$ can 
unambiguously determine $a$ by means of a projective measurement onto the $k$-th 
basis. In fact, he will get the value 
	\beq \label{bval}
b = t \ominus a \, ,
	\eeq
from which, knowing $t$, he can extract $a$.


\subsection{Control Strategy}

Here, we propose an innovative way of realizing the control process to guarantee 
the security of the protocol. Instead of the usual quantum measurement \cite{EM09}, 
we introduce the control by means of a unitary transformation applied by Alice.
Such an operator should realize a permutation of vectors within each basis, to 
allow Bob a reliable data gathering, but not cyclic shift, to differ from the
encoding.

A unitary transformation $W$, satisfying such conditions, can be defined as acting 
on the computational basis in the following way: 
	\beq
W \, \ket{v^0_t}
= \ket{v^0_{\ominus t}} \, .
	\eeq

Then, for each other basis $k$, with $k \neq 0$, we have:
	\beq
W \, \ket{v^k_t}
= {1 \over \sqrt d} \sum_{q=0}^{d-1}
\omega^{\ominus q \odot t}
(\omega^{(k-1) \odot q \odot q})^{\frac{1}{2}} W \, \ket{v^0_q}
= {1 \over \sqrt d} \sum_{s=0}^{d-1}
\omega^{s \odot t}
(\omega^{(k-1) \odot (\ominus s) \odot (\ominus s)})^{\frac{1}{2}} \ket{v^0_s}
= \ket{v^k_{\ominus t}} \, .
	\eeq

That is, $W$ performs the Galois field opposite for each basis ($k=0, \ldots, d$) 
as follows:
	\beq
W \, \ket{v^k_t}
= \ket{v^k_{\ominus t}} \, .
	\eeq

Notice that this transformation satisfies the condition above indicated, only 
when $d$ is an odd prime power dimension.
In fact, for $d = 2^m$ the $W$ operator reduces to the identity , which is not 
acceptable.
It seems reasonable to suppose that it does not exist any transformation of this
kind when $d$ is a power of 2, and moreover that $W$ is the only kind of operator 
with the required properties when $d$ is an odd prime.


\subsection{Description of the protocol}

Then, the protocol runs as follows:
\begin{itemize}
	\item[1.] 
	Bob randomly prepares one of the $d^2$ qudit states $\ket{v_{t}^{k}}$,
	with $k = 1, \ldots, d$ and $t = 0, \ldots, d-1$, and sends it to Alice.
	\item[2.] 
	Alice, upon receiving the qudit state has two options.
		\begin{itemize}
			\item[a)] 
			 With probability $c \neq 0$, she performs a control by applying 
			 the unitary operator $W$ (\textit{Control Mode}). 
			 She then sends back to Bob the resulting state. 
			\item[b)] 
			  With probability $1- c$, she encodes a symbol $a \in A$ by applying 
			  the unitary operator $V_0^{a}$ (\textit{Message Mode}).
			  She then sends back to Bob the resulting state. 
		\end{itemize}
	\item[3.] 
	Bob, upon receiving back the qudit state, performs a measurement by 
	projecting over the basis to which the qudit state initially belonged.
	\item[4.] 
	At the end of the transmission, Alice publicly declares on which runs she 
	performed the control mode and on which others the message mode.
	It is important to remark at this point that Alice does not announce the 
	bases because she did not perfom any measurement.
	For noiseless channel and no eavesdropping, Bob will have obtained the qudit 
	resultant from the action of $W$ operator in the control mode runs, while 
	he will have got the encoded symbol $a$ in the message mode runs.
	
\end{itemize}
%


\section{Security of the protocol}

At first, we consider the most elementary of individual attacks: 
the \emph{Intercept-Resend}.
Suppose Eve, to learn Alice's operation, performs projective measurements on 
both paths of the travelling qudit, randomly choosing the measuring basis. 
She will steal the whole information for each message mode run, indipedently 
from the chosen basis. 

However, in each control mode run, she can guess the correct basis (the same of 
Bob) with probability $1/d$, and in this case she is not detected at all. 
If otherwise Eve chooses the wrong basis, which happens with probability $(d-1)/d$, 
she still has a probability $1/d$ to evade detection. 
The last is exactly the probability that a vector belonging to the wrong basis 
by chance will be projected back to the correct vector of the original basis by 
Bob's measurement.
Then, this means that Alice and Bob reveals Eve with probability $(d-1)^2/d^2$,
which is greater than the result found in \cite{EM09}.

\begin{figure}[htbp] 
\begin{center}
\includegraphics[scale=1]{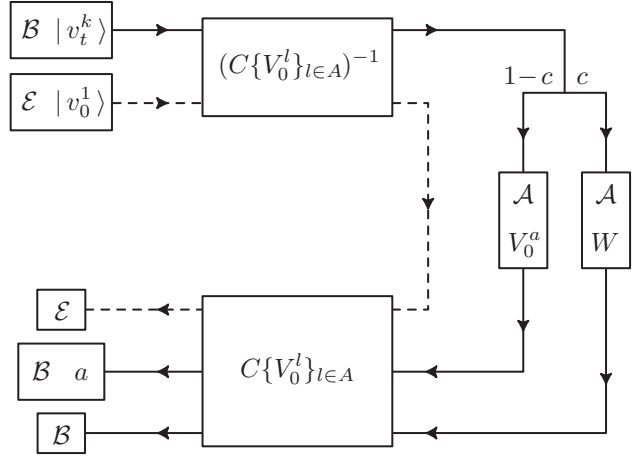}
\end{center}
\vskip-12pt
\caption{The scheme summarizing our protocol. Labels $\mathcal{B}$ and $\mathcal{E}$ 
stand for Bob's and Eve's qudit systems respectively. 
Label $\mathcal{A}$ denotes Alice's operation on Bob's qudit. 
$(C\{V^l_0\}_{l\in A})^{-1}$ and $C\{V^l_0\}_{l\in A}$ represent the eavesdropping 
operations on the forward and backward path respectively.}
 \label{protocol_schemeW} 
	\end{figure}

Now, we are going to evaluate the security of the protocol against a more powerful 
individual attack, already discussed in \cite{EM09}. 
It is known that, quite generally, in individual attacks Eve lets the carrier of 
information interact with an ancilla system she has prepared and then try to gain 
information by measuring the ancilla. 
In this protocol, she has to do that two times, in the forward path (to gain information 
about the state Bob sends to Alice) and in the backward path (to gain information about 
the state Alice sends back to Bob, hence about Alice's transformation).
Moreover, by using the same ancilla in the forward and backward path, Eve could 
benefit from quantum interference effects (see Fig.~\ref{protocol_schemeW}).

As proposed in \cite{EM09}, the attack is described as controlled shifts 
$C\{V^l_0\}_{l\in A}:\mathcal{H}_{d}\otimes\mathcal{H}_{d}
\to\mathcal{H}_{d}\otimes\mathcal{H}_{d}$,
where the controller is the traveling qudit while the target is in the Eve's hands, 
and it is defined as follows:  
	\beq \label{CV^a_0}
\ket{v^1_{t_1}} \ket{v^1_{t_2}} \, \arrow{C\{V^l_0\}_{l\in A}} \,
\ket{v^1_{t_1}} V^{l=t_1}_0 \ket{v^1_{t_2}}
= \ket{v^1_{t_1}} \ket{v^1_{{t_2} \ominus {t_1}}} \, .
	\eeq

We remark that, in this definition, the controller as well as the target states 
are considered in the dual basis for the sake of simplicity. 
Other choices (except the computational basis) will give the same final results.

Then, we consider Eve intervening in the forward path with $(C\{V^l_0\}_{l\in A})^{-1}$, 
defined by 
	\beq
\ket{v^1_{t_1}}\ket{v^1_{t_2}} \, \arrow{(C\{V^l_0\}_{l\in A})^{-1}} \,
\ket{v^1_{t_1}} V^{\ominus {t_1}}_0 \ket{v^1_{t_2}}
= \ket{v^1_{t_1}} \ket{v^1_{{t_2} \ominus ({\ominus t_1})}}
= \ket{v^1_{t_1}} \ket{v^1_{{t_2} \oplus {t_1}}} \, ,
	\eeq
and with $C\{V^l_0\}_{l\in A}$ in the backward path. 

\medskip
%


\subsection{Message Mode}

Now, let us analyze in detail the transformations of the quantum states 
on an entire message mode run.

\bigskip

\noindent {\it Attack on the forward path.}

\medskip

The initial Bob state is one of the $d^{2}$ states $\ket{v_{t}^{k}}$, with 
$k = 1, \ldots, d$ and $t = 0, \ldots, d-1$. 
Then, Eve initially prepares the ancilla state $\ket{v_{0}^{1}}_{\mathcal E}$ 
in the dual basis and performs the controlled operation. Hence, we get
	\beq \label{1attack}
\ket{v^k_t}_{\mathcal B} \ket{v^1_0}_{\mathcal E} 
\, \arrow{(C\{V^l_0\}_{l\in A})^{-1}} \,
\sum_{h=0}^{d-1}
\bracket{v^1_h}{v^k_t}
\ket{v^1_h}_{\mathcal B} \ket{v^1_0}_{\mathcal E} =
\sum_{h=0}^{d-1}
\bracket{v^1_h}{v^k_t}
\ket{v^1_h}_{\mathcal B} \ket{v^1_h}_{\mathcal E} \, .
	\eeq

\newpage

\bigskip

\noindent {\it Encoding.}

\medskip

The Bob's qudit state undergoes the shift $V_0^a$ with $a \in A$, 
then from (\ref{1attack}) we get
	\beq \label{cod}
\, \arrow{\, V^a_0 \,} \, \sum_{h=0}^{d-1}
\bracket{v^1_h}{v^k_t}
\ket{v^1_{h \ominus a}}_{\mathcal B}
\ket{v^1_h}_{\mathcal E} \, .
	\eeq

\bigskip

\noindent {\it Attack on the backward path.}

\medskip

The state (\ref{cod}) undergoes a $C\{V^l_0\}_{l\in A}$ operation, 
hence we have
	\beq \label{2attack}
\, \arrow{C\{V^l_0\}_{l\in A}} \, \sum_{h=0}^{d-1}
\bracket{v^1_h}{v^k_t}
\ket{v^1_{h \ominus a}}_{\mathcal B}
\ket{v^1_{h \ominus (h \ominus a)}}_{\mathcal E}
= \sum_{h=0}^{d-1}
\bracket{v^1_h}{v^k_t}
\ket{v^1_{h \ominus a}}_{\mathcal B} \ket{v^1_a}_{\mathcal E} 
= \ket{v^k_{t \ominus a}}_{\mathcal B} \ket{v^1_a}_{\mathcal E} \,.
	\eeq

Finally, Eve measures her ancilla system by projecting in the dual basis, according 
to the chosen initial ancilla state.

We notice that the controlled operations performed by Eve, as well as her final 
measurement, left unchanged Bob's qudit state. 
Hence, Bob's measurement by projection in the $k$-th basis to which the initial 
state belonged, always allows him to obtain the symbol $a$ Alice has encoded 
[see (\ref{bval})].

On the other hand, Eve gets $\ket{v^1_a}$ with probability 1 as the result of her 
measurement. Therefore, she is able to exactly determine the encoded symbol $a$ 
as well and she steals the whole information, quantified in bits, 
	\beq \label{I_E}
I_\mathcal{E} = \log_2{d} \, ,
	\eeq
on each message mode run.


\subsection{Control Mode}

We would like to evaluate the probability $P_{\mathcal E}$ Alice and Bob have 
to reveal Eve on each control mode run. 
The situation is different for $k=1$ and $k \neq 1$, due to the Eve's choice
of using the dual basis for her ancilla. 
\begin{description}

\item[1)] 
For $k=1$, on the forward path with probability $1/d$ we have:
	\beq
\ket{v_t^{1}}_{\mathcal B}\ket{v^1_0}_{\mathcal E} 
\, \arrow{(C\{V^l_0\}_{l\in A})^{-1}} \,
\ket{v^1_t}_{\mathcal B} \ket{v^1_t}_{\mathcal E} \, .
	\eeq

Then, Alice applies her control strategy:
	\beq
\ket{v^1_t}_{\mathcal B} \ket{v^1_t}_{\mathcal E}	
\, \arrow{\, W \,}	\,
\ket{v^1_{\ominus t}}_{\mathcal B} \ket{v^1_t}_{\mathcal E} \, .
	\eeq

On the backward path it happens the following:
	\beq
\ket{v^1_{\ominus t}}_{\mathcal B} \ket{v^1_t}_{\mathcal E}
\, \arrow{C\{V^l_0\}_{l\in A}} \,
\ket{v^1_{\ominus t}}_{\mathcal B} \ket{v^1_{t \ominus (\ominus t)}}_{\mathcal E}
= \ket{v^1_{\ominus t}}_{\mathcal B} \ket{v^1_{2t}}_{\mathcal E} \, .
	\eeq

Notice that $t \oplus t = 2 \odot t = 2t$ from $0$ to $p-1$, while
$t \oplus t = 2t \neq 2 \odot t$ from $p$ forward being $2 < p$.

It results that Eve's attack does not alter the Bob's and Alice's vectors, hence
Bob, upon his final measurement, will get $\ominus{t}$ with probability 1.
Then, Bob does not outwit Eve's attacks:
	\beq \label{P_E}
P_{\mathcal E} = 0 \, .
	\eeq

\item[2)]
For $k = 2, \ldots,  d$, on the forward path with probability $(d-1)/d$ we get:
	\beq \label{CM_2)1}
\ket{v^k_t}_{\mathcal B} \ket{v^1_0}_{\mathcal E}
= \sum_{h=0}^{d-1}
\bracket{v^1_h}{v^k_t}
\ket{v^1_h}_{\mathcal B} \ket{v^1_0}_{\mathcal E}
\, \arrow{(C\{V^l_0\}_{l\in A})^{-1}} \,
\sum_{h=0}^{d-1}
\bracket{v^1_h}{v^k_t}
\ket{v^1_h}_{\mathcal B} \ket{v^1_h}_{\mathcal E} \, .
	\eeq

Then, Alice applies her control strategy:
	\beq
\sum_{h=0}^{d-1}
\bracket{v^1_h}{v^k_t}
\ket{v^1_h}_{\mathcal B} \ket{v^1_h}_{\mathcal E}
\, \arrow{\, W \,} \,
\sum_{h=0}^{d-1}
\bracket{v^1_h}{v^k_t}
\ket{v^1_\ominus h}_{\mathcal B} 
\ket{v^1_h}_{\mathcal E} \, .
	\eeq

On the backward path it happens the following:
	\beq
\sum_{h=0}^{d-1}
\bracket{v^1_h}{v^k_t}
\ket{v^1_\ominus h}_{\mathcal B} 
\ket{v^1_h}_{\mathcal E}
\, \arrow{C\{V^l_0\}_{l\in A}} \,
\sum_{h=0}^{d-1}\bracket{v^1_h}{v^k_t}
\ket{v^1_\ominus h}_{\mathcal B} \ket{v^1_{h \ominus (\ominus h)}}_{\mathcal E}
= \sum_{h=0}^{d-1}\bracket{v^1_h}{v^k_t}
\ket{v^1_\ominus h}_{\mathcal B} \ket{v^1_{2h}}_{\mathcal E} \, .
	\eeq

Notice that $\ket{v^1_{2h}}_{\mathcal E} = \ket{v^1_{2 \odot h}}_{\mathcal E}$ 
for $2<p$, that is in $G = \mathbb{F}(p^m)$ of characteristic $p > 2$.

In conclusion, the index in the sum is also present in the Eve's ancilla, so the 
Bob's and Eve's states result entangled. Then,
	\beq \label{P_E}
P_{\mathcal E} = \frac{d-1}{d} \,.
	\eeq
\end{description}

In summary, from the two above analized cases, we conclude that the probability 
for Alice and Bob to outwit Eve on each control mode run is
	\beq \label{P_E}
P_{\mathcal E} = 
\left(\frac{1}{d}\right) \cdot 0 + \left(\frac{d-1}{d}\right) \cdot \frac{d-1}{d} = 
\frac{(d-1)^{2}}{d^{2}} \,,
	\eeq
where 
\begin{itemize}

\item $1/d$ is the probability with which Bob and Eve use the same basis (that 
is the dual basis for $k = 1$);

\item $0$ is the corresponding probability of Bob revealing Eve;

\item $(d-1)/d$ is the probability of Eve choosing the basis for ancilla is 
different from Bob's choice of basis for the initial state $\ket{v_t^k}$ 
(then any basis but the dual one, that is $k \neq 1$); 

\item $(d-1)/d$ is anagously the respective probability of Bob outwiting Eve.

\end{itemize}

\medskip
	\begin{figure}[b]
\begin{center}
\includegraphics[scale=0.9]{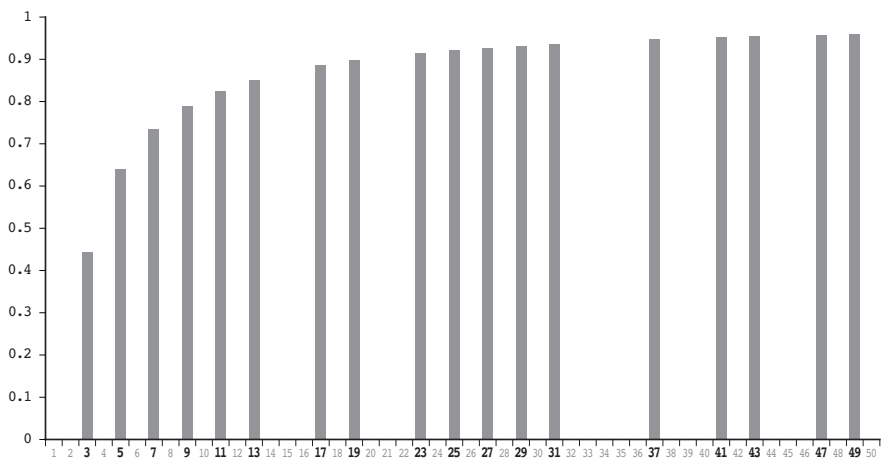}
\end{center}
\vskip-12pt
\caption{The probability $P_{\mathcal E}$ versus the dimension $d$ 
(bars correspond to odd prime power numbers).} 
\label{graf_PeW}
	\end{figure}

Notice that this quantity is largely greater than the analogous obtained with 
control strategy based upon measurement in \cite{EM09}.
Essentially that happens because here the probability $P_{\mathcal E}$ is no 
longer conditioned to the probability that Alice and Bob measure in the same 
bases (this would implicate an other factor $1/d$).
In fact, only Bob perfoms a measurement (at the end of path) and he knowns what 
is the correct basis over which to project (that is the one to which the initial 
qudit state belonged).

The behavior of $P_{\mathcal E}$ as fuction of the order $d$ of the alphabet is 
shown in Fig.~\ref{graf_PeW}.
It can see that the probability $P_{\mathcal E}$ of revealing Eve in each successful 
control mode run increases towards 1 by increasing the dimension $d$. 
Thus, the efficiency of the whole control process increases accordingly to it.


\section{Concluding remarks}

In this paper, we have rivisited the deterministic cryptographic protocol of 
\cite{EM09} which represents a generalization to a $d$-ary alphabet of the 
bidirectional quantum cryptographic scheme of \cite{CL04, LM05, LM05new}.
Here we have introduced a control strategy based on a suitable unitary 
transformation rather than quantum measurements. 
The latter gave an optimal $d = 3$ for the security. 
Now it results that the quantity of information that Eve can steal is the same 
as \cite{EM09}, but the probability $P_{\mathcal E}$ to outwit Eve increases 
in terms of the alphabet order $d$, that is the larger is the alphabet the 
higher is the security.

As a consequence of the deterministic nature of the protocol, this can be also 
used for Quantum Direct Communcation (QDC) between legitimate users 
\cite{BF02, DLL03, DL04, CL04dir, LM05, EM09}, that is when Alice and Bob (after 
authentication) communicate directly the meaningful message without encryption. 
Notice that for this kind of communication only an asymptotic security can be 
proven.
 
Hence, if we assume that Eve wants to perform her attack on each message mode 
run, without having been detected in the previous control mode runs, then the 
probability is given by following geometric series:
\beq
 (1-c) + c(1-P_{\mathcal E})(1-c) + 
c^{2}(1-P_{\mathcal E})^{2}(1-c) + \ldots 
=\frac{1-c}{1-c(1-P_{\mathcal E})} \,.
\eeq

Thus, being $I_{\mathcal E}$ the quantity of information that Eve eavesdrops in 
a single attack, the probability that she successfully eavesdrops an amount of 
information $I$ is
\beq \label{fmla_QDC}
\left(\frac{1-c}{1-c\big(1-P_{\mathcal E}\big)}\right)
^{\!I/I_{\mathcal E}} \,,
\eeq
with $I_{\mathcal E}$ and $P_{\mathcal E}$ given in (\ref{I_E}) and (\ref{P_E}) 
respectively. 

\begin{figure}[htbp]
\begin{center}
\includegraphics[scale=0.9]{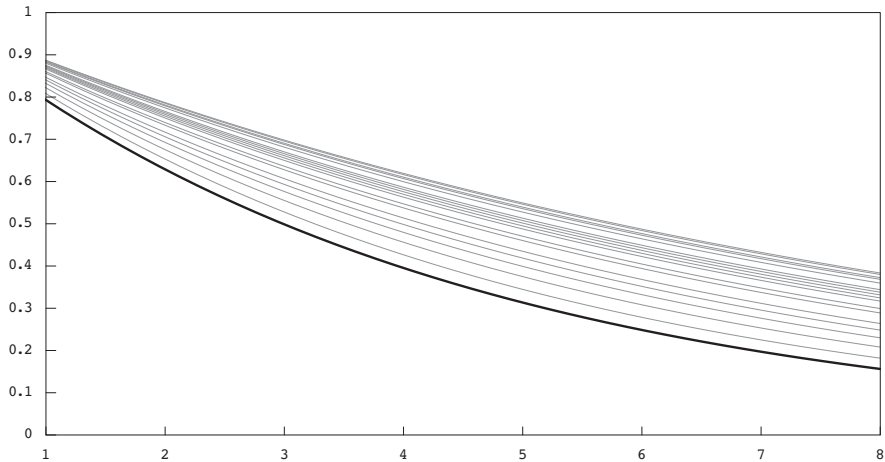}
\end{center}
\vskip-12pt
\caption{The eavesdropping success probability as a function of the maximal 
eavesdropped information decreases faster by increaing $d$. It is plotted for 
different dimensions, from bottom to top 
$d=3, ~d=5, ~d=7, ~d=9, ~d=11, ~\ldots, ~d=49$.}
\label{graf_QDC}
\end{figure}

In Fig.~\ref{graf_QDC} we have plotted the quantity of (\ref{fmla_QDC}), with 
$c = 1/2$, versus the number $n$ of bits stolen by Eve without being outwitted 
for different alphabet order.
It is interesting to observe that such a probability, as a function of $I$,
increases slowly and slowly with the alphabet order.

\medskip

In this case the probability for Alice and Bob to detect Eve before she can 
eavesdrop a fixed amount of information, that is the complement of probability 
in (\ref{fmla_QDC}), is maximal for $d=3$. 
Notice that the optimal dimension depends on the specific task of the protocol 
(QKD or QDC). 

We believe that this work might offer new interesting perspectives for deterministic 
cryptographic protocols, in particular it could stimulate further studies about the 
optimal control strategy.


\acknowledgments

We have the pleasure of thanking R. Piergallini for several and stimulating 
discussions on this subject.




\begin{thebibliography}{99}

%
\bibitem{BB84}
C. H. Bennett and G. Brassard,  
Proc. of IEEE Int. Conf. on Computers, Systems and Signal Processing, 
Bangalore, India (IEEE, New York, 1984).

%
\bibitem{BEKW02} 
A. Beige, B.-G. Englert, C. Kurtsiefer, H. Weinfurter, 
J. Phys. A: Math. Gen. \textbf{35}, L407 (2002).

%
\bibitem{BF02} 
K. Bostr\"{o}m and T. Felbinger,
Phys. Rev. Lett. \textbf{89}, 187902 (2002).

%
\bibitem{CL04}
Q.-Y. Cai and B.-W. Li, 
Chin. Phys. Lett. \textbf{21}, 601 (2004).

%
\bibitem{LM05} 
M. Lucamarini and S. Mancini, 
Phys. Rev. Lett. \textbf{94}, 140501 (2005).

%
\bibitem{LM05new} 
M. Lucamarini and S. Mancini, 
arXiv:1004.0157.

%
\bibitem{BT99}
H. Bechmann-Pasquinucci and W. Tittel,
Phys. Rev. A \textbf{61}, 062308 (2000).

%
\bibitem{CBKG01}
N. Cerf, M. Bourennane, A. Karlsonn and N. Gisin,
Phys. Rev. Lett. \textbf{88}, 127902 (2002).

%
\bibitem{SLW06}
J. S. Shaari, M. Lucamarini and M. R. B. Wahiddin,
Phys. Lett. A \textbf{358}, 85 (2006);

%
\bibitem{SW07}
J. S. Shaari and M. R. B. Wahiddin,
Phys. Lett. A \textbf{361}, 445 (2007).

%
\bibitem{PMBL06}
S. Pirandola, S. Mancini, S. Braunstein and S. Lloyd,
Nat. Phys. \textbf{4}, 726 (2008).

%
\bibitem{EM09}
A. Eusebi and S. Mancini,
Quantum Inf. \& Comp. \textbf{9}, 950 (2009).

%
\bibitem{DLL03}
F.-G. Deng, G. L. Long and X.-S. Liu, 
Phys. Rev. A \textbf{68}, 042317 (2003).

%
\bibitem{DL04}
F.-G. Deng and G. L. Long, 
Phys. Rev. A \textbf{69}, 052319 (2004).

%
\bibitem{CL04dir}
Q.-Y. Cai and B.-W. Li, 
Phys. Rev. A \textbf{69}, 054301 (2004).

%
\bibitem{I81}
I. D. Ivanovic, J. Phys. A \textbf{14}, 3241 (1981).

%
\bibitem{WF89}
W. K. Wootters and B. D. Fields, Ann. Phys. \textbf{191}, 363 (1989).

%
\bibitem{BBRV01} 
S. Bandyopadhyay, P. O. Boykin, V. Roychowdhuri and F. Vatan, 
Algorithmica \textbf{34}, 512 (2002).

%
\bibitem{KR03}
A. Klappenecker and M. R\"otteler,
Finite Fields and Applications, 137, Lecture Notes in Comput. Sci., 
\textbf{2948}, Springer, Berlin, (2004).

%
\bibitem{D05}
T. Durt, 
J. Phys. A: Math. Gen. \textbf{38}, 5267 (2005).


\end{thebibliography}
\end{document}